\documentstyle[11pt,newpasp,twoside,epsf]{article}
\def\edcomment#1{\iffalse\marginpar{\raggedright\sl#1\/}\else\relax\fi}
\reversemarginpar

\begin{document}
\title{Luminosity and Redshift Dependence of Quasar Spectral Properties}
 \author{Daniel E. Vanden Berk, Ching-Wa Yip, Andrew J. Connolly}
\affil{Univ. of Pittsburgh, Dept. of Physics and Astronomy, 3941 O'Hara St.,
  Pittsburgh, PA, 15260} 
\author{Sebastian Jester, Chris Stoughton}
\affil{Fermilab, 500 Wilson Rd., Batavia, IL, 60510}

\begin{abstract}
Using a large sample of quasar spectra from the SDSS, we examine the
composite spectral trends of quasars as functions of both redshift
and luminosity, independently of one another.  Aside from the well
known Baldwin effect (BE) -- the decrease of line equivalent width
with luminosity -- the average spectral properties are remarkably
similar.  Host galaxy contamination and the BE are the primary causes
for apparent changes in the average spectral slope of the quasars.
The BE is detected for most emission lines, including the Balmer
lines, but with several exceptions including NV1240A.  Emission line
shifts of several lines are associated with the BE.  The BE is mainly
a function of luminosity, but also partly a function of redshift in
that line equivalent widths become stronger with redshift.  Some of
the complex iron features change with redshift, particularly near
the small blue bump region.
\end{abstract}

\section{Introduction}
The Sloan Digital Sky Survey (SDSS, York et~al.\ 2000) is now identifying
tens of thosands of new quasars a year.  The large sample, wide ranges
of both redshift and luminosity, and the high-quality calibrated spectra,
make the SDSS quasar sample extraordinarily useful for exploring the
dependence of spectral properties on redshift and luminosity -- two of
the most important parameters for any extragalactic population.  Here
we present initial results on the composite spectral properties of
more than 16000 quasars from the SDSS.

\begin{figure}
\plotfiddle{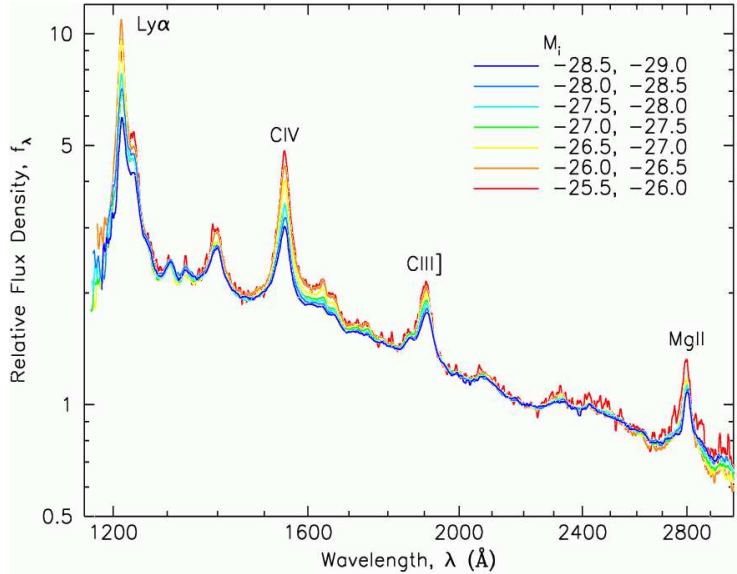}{2.7in}{-90}{40}{40}{-150}{225}
\caption{Composite quasar spectra from the same small range of redshift
  but at different luminosities.  The flux densities have been normalized
  at 2200{\AA}.  The spectra are nearly indistinguishable except for the
  Baldwin effect -- the decrease in emission line equivalent width with
  luminosity.}
\end{figure}

\section{The Dataset and Composite Spectrum Construction}
The quasar data are taken mainly from the SDSS First Data Release
(DR1, Abazajian et al.\ 2003), appended with several hundred more
post-DR1 quasar spectra in order to extend the redshift and luminosity
coverage.  Quasars in the DR1 are described by (Schneider et al.\ 2003).
We modify the definition of quasar here to mean any extragalactic
object with at least one emission line FWHM of at least $1000$km/s,
and impose no luminosity criteria.  We also remove from consideration
any spectrum with an apparent broad absorpion line or strong
associated absorption line, since these significantly affect the
true profiles of some emission lines.  Spectra with long unprocessable
wavelength ranges are also rejected.  The luminosity is measured as
the rest frame K-corrected $i$ band absolute magnitude, $M_i$, measured at
the spectral epoch to avoid variability effects, using a flat cosmology
with $\Omega_{M} = 0.3, \Omega_{\Lambda}=0.7, H_{0}=70$km/s/Mpc.

The full sample consists of 16716 independent quasar spectra from the SDSS.
The sample was divided into redshift bins of width $\Delta log(1+z)=0.04$
starting at $z=0$, and absolute magnitude bins of half a magnitude.
Composite spectra were generated for each bin by calculating the geometric
mean of all of the spectra contained in the bin, using techniques similar
to those described by Vanden Berk et~al.\ (2001).  The geometric mean
preserves the average index of a set of power laws.  We have found that
the resulting composite spectra do not change greatly when at least about 
20 spectra are combined.  In this analysis we consider only bins with at
least 20 spectra; there are 133 such bins, spanning redshifts beyond 4.7
and absolute magnitudes from -20 to -29.5.

\section{Results}
One of the primary conclusions of this study is that the {\em average}
spectral properties of quasars do not vary greatly as a function of either 
redshift or luminosity.  The primary trend with either parameter is the
decrease of most emission line equivalent widths with luminosity -- the
well-known Baldwin effect (BE, Baldwin 1977).  To illustrate this, Fig.\,1
shows seven composite spectra from the same redshift bin,
$2.020 \le z < 2.311$, but spanning more than an order of magnitude in
luminosity.  The spectra are normalized to unity at a rest wavelenth
of 2200{\AA}, and they are color coded to show the absolute magnitue bin
range.  The BE is obvious for most emission lines, but especially
Ly\,$\alpha$ and C{\sc iv}\,$\lambda$1549. Notable exceptions include
O{\sc I}, Al{\sc iii}, and most likely N{\sc v}.

\begin{figure}
%\plotfiddle{zedColSpec2.020.eps}{3in}{-90}{50}{50}{0}{100}
%\plotone{c4be.eps}
\plottwo{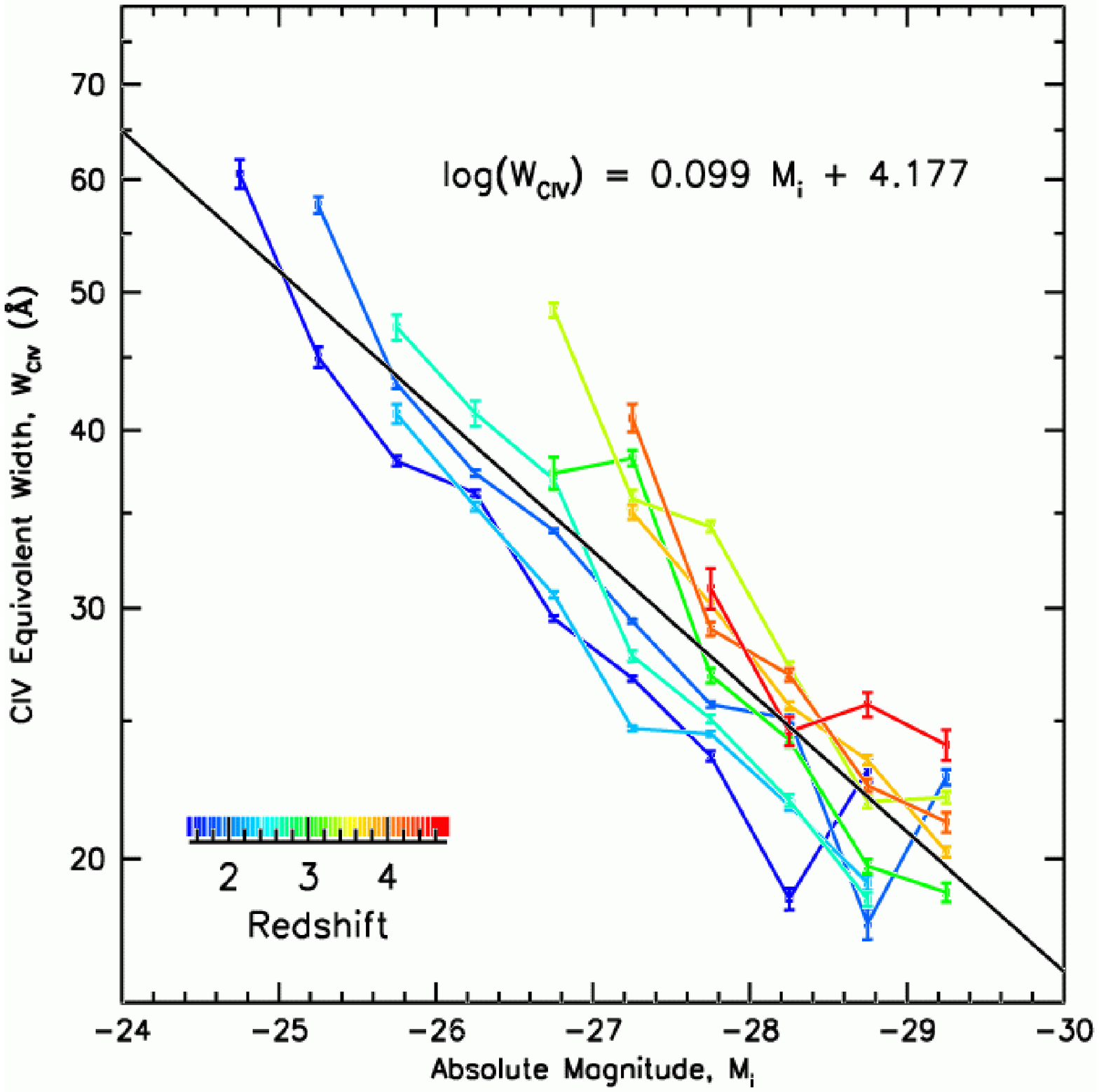}{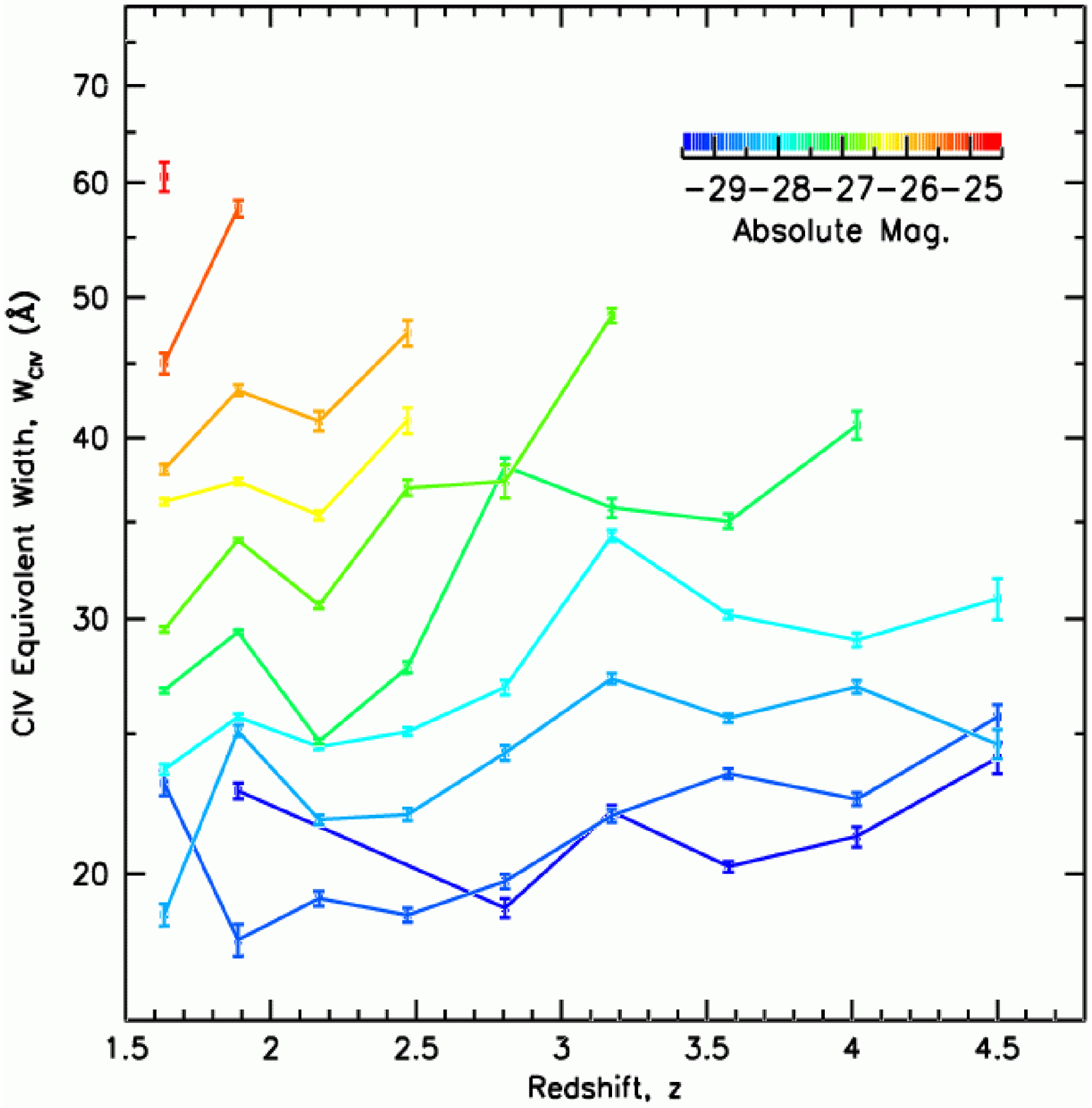}
\caption{The C{\sc iv} equivalent width as a function of both luminosity
  (left) and redshift (right) in small ranges of the other parameter.}
\end{figure}

It can be discerned from Fig.\,1 that the peaks of several of the lower
ionization lines, e.g. Mg{\sc ii}\,$\lambda$2800, shift redward with 
increasing luminosity.  Since these lines are expected to lie at close
to the systemic redshifts of the quasars, and the measured redshifts
rely more strongly on the C{\sc iv} high-ionization line, in reality
it is the high-ionization lines that are increasingly {\em blushifted}
with luminosity.  This confirms the relationship between line shift
and the Baldwin effect found by Richards et~al.\ (2002).

The continuum slopes of the spectra in Fig.\,1 are remarkably similar.
However, at redshifts below about 0.5, the slopes become bluer with
luminosity.  The reason for this is that host galaxy contamination 
increases with decreasing luminosity, making the quasar spectrum appear
redder.  We have confirmed this by fitting the low-$z$ composite spectra
with a pure quasar and a pure galaxy component using sets of eigenvectors.
After subtracting the host galaxy component, the quasar spectra reveal
a Baldwin effect for the Balmer lines and the narrow forbidden lines,
a point which has previously been controversial.

While the Baldwin effect is a strong function of luminosity, Fig.\,2
shows that it is also a function of redshift.  The equivalent width of
C{\sc iv} is shown as a function of luminosity and redshift, in bins
in which the other parameter is held nearly constant.  The redshift
dependence is weaker, but still quite significant.  The redshift
dependence may explain much of the ``scatter'' that is often found
in the Baldwin effect; unfortunately this also means that the BE will
remain difficult to use for cosmology since it evolves with redshift.

Finally, there are other redshift trends to note, mainly that the
strengths and profiles of many of the iron complexes evolve.  This
occurs most notably in the so-called small blue bump region -- the
entire region from about 2200-4000{\AA} appears to become bluer with
decreasing redshift.  Some of this may be due to changes in the
Balmer continuum, but the Fe emission complex profiles also change,
indicating that Fe itself contributes at least partially to the effect.

%% Acknowledge SDSS
\vskip 3ex
Funding for the Sloan Digital Sky Survey (SDSS) has been provided
by the Alfred P. Sloan Foundation, the Participating Institutions,
the National Aeronautics and Space Administration, the National
Science Foundation, the U.S. Department of Energy, the Japanese
Monbukagakusho, and the Max Planck Society. Participating
Institutions

The SDSS is a joint project of The University of Chicago, Fermilab,
the Institute for Advanced Study, the Japan Participation Group,
The Johns Hopkins University, Los Alamos National Laboratory, the
Max-Planck-Institute for Astronomy (MPIA), the Max-Planck-Institute
for Astrophysics (MPA), New Mexico State University, University
of Pittsburgh, Princeton University, the United States Naval
Observatory, and the University of Washington.

%% References

\end{document}